\pdfoutput=1
\documentclass[10pt, conference]{IEEEtran}

\usepackage{cite}

\ifCLASSINFOpdf
   \usepackage[pdftex]{graphicx}
   \usepackage{epstopdf}
  \graphicspath{{../pdf/}{../jpeg/}}
  \DeclareGraphicsExtensions{.pdf,.jpeg,.png,.eps}
\else
   \usepackage[dvips]{graphicx}
  \graphicspath{{../eps/}}
  \DeclareGraphicsExtensions{.eps}
\fi

\usepackage{color,soul}

\usepackage{amsmath}
\usepackage{amssymb}
\usepackage{amsthm}
\usepackage{algorithm}
\usepackage{algpseudocode}

\ifCLASSOPTIONcompsoc
  \usepackage[caption=false,font=normalsize,labelfont=sf,textfont=sf]{subfig}
\else
  \usepackage[caption=false,font=footnotesize]{subfig}
\fi

\usepackage{balance}

\usepackage{cite}
\usepackage{amsmath,amssymb,amsfonts}
\usepackage{algorithm}
\usepackage{algpseudocode}
\usepackage{graphicx}
\usepackage{textcomp}
\usepackage{xcolor}
\usepackage{verbatim}
\usepackage[utf8]{inputenc}
\usepackage{bm}

\newtheorem{theorem}{Theorem}

\usepackage[T1]{fontenc}

\begin{document}


\title{Multi-Entanglement Routing Design  over \\ Quantum Networks }

\author{\IEEEauthorblockN{
Yiming Zeng\IEEEauthorrefmark{3},
		Jiarui Zhang\IEEEauthorrefmark{3},
		Ji Liu,
		Zhenhua Liu,
		Yuanyuan Yang
}
	\IEEEauthorblockA{
		Stony Brook University, Stony Brook, NY 11794, USA}
	\{yiming.zeng, jiarui.zhang.2, ji.liu, zhenhua.liu, yuanyuan.yang\}@stonybook.edu
}

\IEEEoverridecommandlockouts
\IEEEpubid{\makebox[\columnwidth]{
\IEEEauthorrefmark{3} Both authors contributed equally to this research. \hfill
} \hspace{\columnsep}\makebox[\columnwidth]{ }}

\maketitle

\begin{abstract}
Quantum networks are considered as a promising future platform for quantum information exchange and quantum applications, which have capabilities far beyond the traditional communication networks. Remote quantum entanglement is an essential component of a quantum network. How to efficiently design a multi-routing entanglement protocol is a fundamental yet challenging problem. In this paper, we study a quantum entanglement routing problem to simultaneously maximize the number of quantum-user pairs and their expected throughput. 
Our approach is to formulate the problem as two sequential integer programming steps. We propose efficient entanglement routing algorithms for the two integer programming steps and analyze their time complexity and performance bounds. Results of evaluation highlight that our approach outperforms existing solutions in both served quantum-user pairs numbers  and the network expected throughput.  

\end{abstract}


\section{Introduction}
Quantum networks enable to generate, transmit and compute quantum information (qubits) in addition to classical data between quantum (ebits) processors~\cite{van2014quantum}. 
It supports massive quantum applications in both quantum computing and quantum communication systems, such as distributed quantum computing~\cite{cacciapuoti2019quantum,caleffi2018quantum}, quantum communication~\cite{gisin2007quantum}, quantum machine learning~\cite{biamonte2017quantum} and quantum key distribution\cite{scarani2009security}. 
Several quantum systems have been constructed, such as long-distance link (40 kilometers) teleportation over the fiber link~\cite{valivarthi2020teleportation}, the mobile quantum network~\cite{liu2021optical}, and the integrated entanglement system through satellites which can support the entanglement over 4600 kilometers~\cite{chen2021integrated}. 

Entanglement is an essential component of most quantum applications mentioned above. 
For example, the quantum key distribution system has  provable security for the distributed information~\cite{van2014quantum} by taking advantage of the entanglement and no-cloning theorem~\cite{wootters2009no}.  
Supporting the long-distance entanglement is  fundamental  for quantum networks. 
However, the entanglement process is probabilistic and not stable. 
Different from binary ebits in  traditional communication, qubits created by photons are extremely fragile.
The successful entanglement rate among qubits decreases exponentially with the transmission length. 
Hence, to enable long-distance entanglement of quantum users in the quantum network, quantum switches are placed in the network as relays to supply end-to-end entanglements for multiple quantum users that demand them~\cite{van2013designing, vardoyan2019stochastic}. 
Quantum switches are equipped with quantum memories (qubits) and have the ability to perform multi-qubits measurement (swapping)~\cite{vardoyan2019stochastic}. 

The \textit{entanglement routing} problem about
\textit{how to build long-distance entanglement through quantum switches}  is crucial in the quantum network.  
Thoughtful design for the entanglement routing in the quantum network can boost the network performance  by efficiently utilizing resources, e.g., switch memories. 

While large-scale quantum networks have not been implemented out of the lab due to physical and experimental challenges, it is still valuable to investigate the  entanglement routing problem from the network layer for the future. 
The entanglement routing problem has been drawing great attention in previous studies. 
\cite{pant2019routing,li2021effective,chakraborty2019distributed,vardoyan2019stochastic,shchukin2019waiting,das2018robust} study the entanglement routing problem or theoretical entanglement performance on the special network topologies such as a single switch, single entanglement path, rings, grids, or spheres. 
\cite{shi2020concurrent,zhang2021fragFmentation} consider a general quantum network for multiple quantum users pairs entanglement. 
However, their strategy is a greedy algorithm to maximize the throughput of the quantum user pair one by one which might assign too many resources to limited quantum users, and other quantum users are neglected. 
The proposed algorithm incurs high time complexity and lacks the performance guarantee. 

Moreover, most existing works treat the transmission link capacity as the main bottleneck of the network.
However, the switch resource (the number of qubits) is the limitation of the quantum network in reality instead of the transmission link capacity. 
A most recent quantum processor can only have up to 8 qubits~\cite{dahlberg2019link}. 
An optical fiber cable can contain up to 25 cores, each core can be used as an independent link for the entanglement.
Multiple optical fiber cables can be placed between quantum  switches. 
Hence, the transmission link has enough capacity to serve the entanglement demands for the quantum users in current quantum networks. 

In this paper, we consider a general quantum network structure and present a comprehensive entanglement process for multiple pairs of quantum users. 
Our goal is to \textit{maximize the number of quantum-user pairs and the expected network throughput at the same time}. 
Our contributions are as follows:
\begin{enumerate}
    \item We describe the detailed  multi-entanglement routing  process  for multiple quantum-user pairs as the offline and the online stages.  
    \item We formulate the problem as two integer linear programming problems that are both NP-Complete.
    \item We design the routing protocol by proposing efficient algorithms with lower time complexity and performance guarantees. 
    \item Results of evaluation highlight that our approach can improve the number of served quantum-user pairs 85\% and the expected throughput 27\% in average compared with existing works.  
\end{enumerate}
To the best of our knowledge, this is the first paper to maximize the network served quantum-user pairs number and expected throughput simultaneously. 

The organization of the paper is as follows: we first introduce the background of the quantum network and the multi-entanglement routing process in Section~\ref{sec: background}. 
Then, we present the quantum network model and formulate the routing entanglement process as two integer linear programming problems in Section~\ref{sec: network model}  based on the routing process introduced  in Section~\ref{sec: background}. 
The entanglement routing algorithms are proposed in Section~\ref{sec:Alg} and Section~\ref{sec:alg2} for two integer linear programming problems, respectively. 
We conduct extensive simulations to discuss and analyze the performance of our proposed algorithms  and compare them with previous work in Section~\ref{sec:simulation}, followed by related work in Section~\ref{sec:related-work} 
the conclusion in Section~\ref{sec:conclusion}.

\section{Quantum Network Background} \label{sec: background}
In this section, we introduce some basic quantum network backgrounds, including quantum network components and multi-routing entanglement processes.   

\subsection{Quantum Communication}
\subsubsection{Qubit} In the quantum network or quantum computing, a qubit is the basic unit to represent quantum information. A qubit can be an electron or a photon or a nucleus from an atom.  A qubit is described by its state~\cite{van2014quantum}. Different from an  ebit in the classical Internet representing 0 or 1, a qubit can present a coherent superposition of both. 

\subsubsection{Entanglement} 
Entanglement is a phenomenon that a group of qubits expresses a high correlation state which can not be explained by individual qubits states.  In this paper, we consider the simplest case of two qubits entanglement which is bipartite entangled states. In quantum physic, a simple way to entangle two independent qubits is by using CNOT gate~\cite{nielsen2002quantum}.  
When the entanglement qubits number is two,  Bell-state measures (BSMs) can be applied to measure the entanglement. 

\subsubsection{Teleportation}
\begin{figure}[htbp]
\vspace{-0.1in}
\centerline{\includegraphics[width=0.25\textwidth]{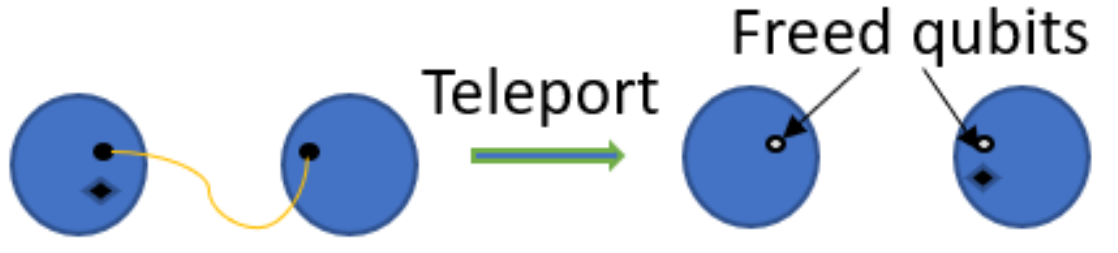}}
\caption{A Teleportation Example. The source node teleport a qubit by a pair of entanglement qubits.   }
\label{fig:teleport}
\vspace{-0.1in}
\end{figure}

If a pair of entanglement quibts are shared by two nodes, the secret information can be transmitted from one node to another one with the help of quantum measurement. This process is called teleportation. An example is illustrated in Figure~\ref{fig:teleport}. 

\subsubsection{Entanglement Swapping} 
\begin{figure}[htbp]
\vspace{-0.1in}
\centerline{\includegraphics[width=0.4\textwidth]{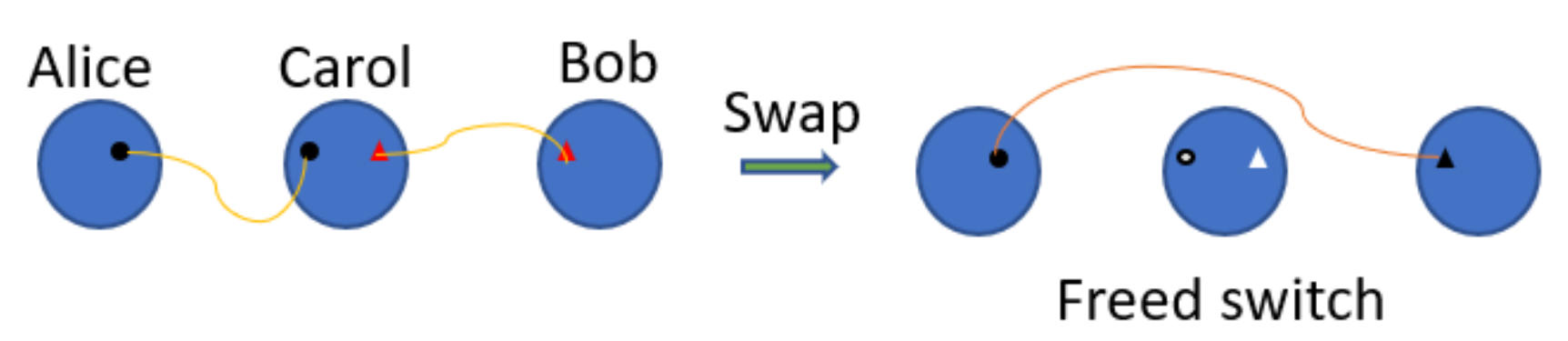}}
\caption{A Swapping Example.}
\label{fig:swap}
\vspace{-0.2in}
\end{figure}

Figure~\ref{fig:swap} presents an example of swapping.
If Alice shares an entangled qubit pair (Bell pair) with the middle node Carol, and Carol shares another entangled qubit pair with Bob, Carol can teleport its qubit entangled with Alice to Bob, then Alice and Bob are entangled directly~\cite{coecke2014logic}. 

\subsection{Quantum Network Components}
With these basic concepts, we first introduce several important components in the quantum network. 

\subsubsection{Quantum Users}
A quantum user has demands to entangle with another user in the quantum network for quantum communication. We assume that there is a direction between a pair of quantum users intending to be entangled. 
The user who intends to entangle with another user is called a source node.  Another user who tried to be entangled is called a destination node.

\subsubsection{Quantum Switches}
The quantum switch is a node with quantum memories to work as relays for the entanglement process in the quantum network~\cite{briegel1998quantum, vardoyan2019stochastic}. 
They can either transmit qubits or establish the entanglement at distant nodes without physically sending an entangled qubit by swapping. 

\subsubsection{Quantum Links}
Quantum links are the links used for connecting quantum switches and quantum users. In this paper, we assume that the quantum network is connected by optical fiber cables among quantum switches and quantum users.
The successful entanglement generation probability is related to the material and the length of the quantum link, i.e., $p=e^{-\alpha L}$, where $\alpha$ is a positive constant related to the material of the quantum link and $L$ is the length of the quantum link. 

\subsubsection{The Traditional Internet (The Cloud)}
The quantum network co-works with the traditional Internet together for quantum users' entanglement routing. 
The Internet is responsible for exchanging information among the networks. 
For each node including quantum users and quantum switches, they are equipped with traditional computing devices  (e.g., computers) and can communicate through the traditional Internet. 

We list several of the most important roles of the traditional Internet (the cloud) in the quantum network, but not including all. 
\begin{itemize}
    \item The cloud is the center of the network that knows detailed information of quantum network including quantum-user pairing information, the quantum network topology, the quantum switch capacity, and so on. 
    \item The cloud computes the offline routing paths of quantum-user pairs with network information available. 
    \item The cloud shares network information through the Internet such as quantum-user pairing information, routing paths to quantum switches. 
    \item During the entanglement process, adjacent switches (e.g., the graph distance between switches is small) communicate through the Internet to inform each other about link and switch states. 
    
\end{itemize}

\subsection{Entanglement Process} \label{sec:ep}
\cite{shi2020concurrent} presents the detailed quantum network entanglement process for one quantum-user pair. 
Here, we summarize the routing entanglement process for multiple quantum-user pairs as a two-stage process including an \textit{offline stage} and an \textit{online stage}.

\subsubsection{Offline Stage}
In \textit{offline stage}, the main tasks of the quantum network are offline entanglement routing design for quantum-user pairs and transmitting the routing paths to switches for the entanglement in \textit{online stage}. 

The offline routing protocol design is conducted by the cloud.  
We assume that the following offline information of the network is known by the cloud: the quantum-user pairing information; the network topology (switches placement and connection); switches information (the number of qubits in each switch). 
With all information available, the cloud computes the routing paths for quantum-user pairs with the limitation of switches capacity, and the detailed computing process is discussed in Section~\ref{sec:Alg} and Section~\ref{sec:alg2}.  
After that, the routing paths computed by the cloud are transmitted through the Internet to switches for the entanglement.  

\subsubsection{Online Stage}
In \textit{online stage}, the switches try to generate entanglement among links with the routing paths sent from the cloud, and then swap in the interiors. 

The entanglement and swapping process is probabilistic, e.g., the successful entanglement rate over an optical fiber is typically 0.01\%~\cite{dahlberg2019link}. 
The duration of the entanglement over a link is short, e.g., 1.46s~\cite{dahlberg2019link}. The entanglement generation time of one attempt is usually  165 $\mu s$~\cite{dahlberg2019link}. 
All the entanglement and swapping processes over a path should be processed in the duration of the entanglement $T$. 
The short duration of $T$ requires the entanglement and swapping process to be carefully considered.

The detailed entanglement process is as follows. 
\begin{itemize}
    \item First, all the switches are time-synchronized through the Internet~\cite{pant2019routing} which can ensure the whole quantum network starts entanglement at the same time. 
    \item Second, all the switches try to process entanglement over links and swap in the interiors given the routing paths of all quantum-user pairs. Each switch can try multiple times until the entanglement is generated or the time out (greater than $T$). 
    \item Third, some switches may fail to generate entanglement over part of links to build a path for quantum-user pairs. 
    Then, the switches will try to build a recovery path for quantum-user pairs locally.  
    Link states (entanglement or not) and swapping states cannot be efficiently sent to the cloud for rescheduling in $T$ due to the Internet delay. 
    The switch can access link states near it through communication with nearby switches with the Internet. The transmission delay from a switch to other switches in a few hops is acceptable compared with $T$. The exact number of hops depends on the Internet latency condition. A typical communication time between two switches within one hop is around 1 $ms$~\cite{dahlberg2019link}. 
    With the link states and swapping states available, the switches decide the recovery paths for quantum-user pairs locally.  
\end{itemize}

In this paper, we focus on the entanglement routing design in \textit{offline stage}. The recovery path design in \textit{online stage} has been addressed in \cite{shi2020concurrent, zhang2021fragFmentation} efficiently in an online manner which can be applied to our model directly.   

\section{Quantum Network Model} \label{sec: network model}
\begin{figure}[htbp]
\centerline{\includegraphics[width=0.49\textwidth]{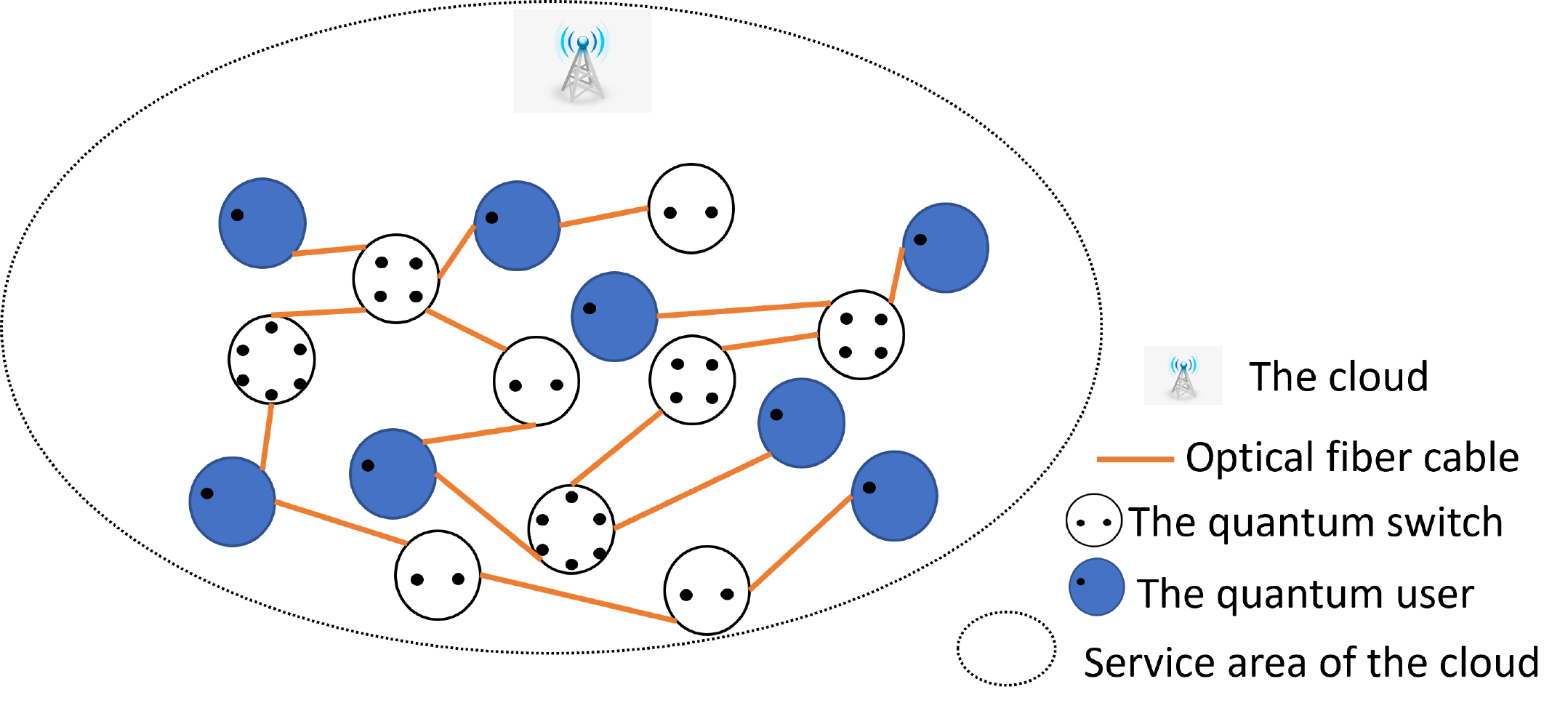}}
\caption{An Example of Network.}
\label{fig:network}
\vspace{-0.2in}
\end{figure}

In this section, we first describe the quantum network model, and then formulate the routing entanglement problem with the goal to maximize the number of  quantum-user pairs that can be served by the network and their expected throughput. 
The network model described here follows real quantum network entanglement experiments~\cite{pan1998experimental,zhang2019experimental,bouwmeester1997experimental} and previous studies about quantum entanglement routing~\cite{shi2020concurrent,van2013designing,pant2019routing,vardoyan2019stochastic}. 
Figure~\ref{fig:network} shows  an  example of the proposed quantum network.
The key notations are summarized in Table~\ref{notation}. 

 \begin{table}[tbp] 
\caption{Table of Notations Used In the  Network Model}
\label{notation}\centering
\begin{tabular}{cl}
\hline
Notation & Definition \\ \hline

$\mathcal{M}$ & The set of $\langle S,D\rangle$ pairs \\
$\mathcal{A}$ & The path set of $\langle S,D\rangle$ pairs \\
$\mathcal{A'}$ & The path set of $\langle S,D \rangle$ pairs with \\
& $M^2$ shortest distance paths \\
$\hat{\mathcal{M}} $ & The set of $\langle S,D\rangle$ pairs selected in \textsc{STEP I} \\
$\mathcal{S}$ & The set of source $\{s_1, s_2, \ldots, s_M\}$ \\
$\mathcal{D}$  & 
The set of destinations $\{d_1, d_2, \ldots, d_M\}$\\
$\mathcal{V}$ & The set of switch nodes  $\{ {v_1,v_2,\ldots,v_N}\}$\\
$\mathcal{E}$ &  The set of connection links between switches  \\
& $\{ e_{ij}, v_i, v_j \in V \}$ \\
$L_{ij}$ & Length of link $e_{ij}$ \\
$Q_i$ & The number of qubits contained by switch $v_i$ \\
$\hat{Q}_i$ & The number of available qubits of switch $v_i$ after \textsc{STEP I} \\
$p_{ij}$ & The successful entanglement rate of edge $e_{}ij$ \\
$\alpha$ & Link transmission efficiency \\ 
$\mathcal{A}_m$ & The set of all paths for $\langle s_m, d_m\rangle$ \\
$A_m$ & A path belongs to $\mathcal{A}_m$  \\ 
$\mathcal{A'}_m$ & The set of $M$ shortest distance paths for $\langle s_m, d_m\rangle$ \\
$A'_m$ & A path belongs to $\mathcal{A'}_m$ \\ 
$\mathcal{P}_{A_m}$ & The expected output qubits of path $A_m$  \\ 
$Q^{A_m}$ &  The number of qubits assigned to path $A_m$ \\ 
$x_{A_m} \in \{0,1\}$ & Binary variable indicates whether path $A_m$ \\ & is selected in the network \\                    
\hline
\end{tabular}
\vspace{-0.2in}
\end{table}
 
\subsection{Network Model}
\textbf{Quantum Users}: The Quantum user set $\mathcal{M}$ consists of  $M$ quantum-user pairs $\langle s_1, d_1\rangle, \langle s_2, d_2\rangle, \cdots, \langle s_M, d_M\rangle$. 
$\mathcal{S}=\{s_1, s_2, \cdots, s_m\}$ denotes the set of sources, and $\mathcal{D}=\{d_1, d_2, \cdots, d_m\}$ denotes the set of destinations. 
They are connected with the quantum network and request for  the entanglement. 
In this paper, we assume that quantum users do not involve  other quantum-user pairs entanglement process as switches, and all switches are honest and controlled by the cloud to serve the network.  

\textbf{Network Graph}: The transmission graph consists of quantum switches and quantum links. 
The network is abstracted as an undirected graph which is denoted as  $G=(\mathcal{V},\mathcal{E})$, where $\mathcal{V}=\{ {v_i}\}_{i=1}^N$ denotes the set of quantum switches, and $\mathcal{E}= \{ e_{ij}\}\subset \{(v_i,v_j)\;: \; v_i, v_j \in \mathcal{V} \}$ denotes the set of the quantum links. 

\textbf{Quantum Switch}: Each quantum switch $v_i \in \mathcal{V}$ has $Q_i$ qubits that can be assigned for the entanglement. 
We focus on 2-qubit entanglement, and the switch uses Bell-state measurements (BSMs). Since exact 2 qubits will be involved in the swapping process,
we assume that $Q_i$ is a positive even number. 
This assumption also fits the real switch design~\cite{dahlberg2019link}. 
The successful swapping rate in each switch 
for any pair of qubits is uniform and denoted as $q \in [0,1]$.

\textbf{Quantum Link}:
$e_{ij}$ is an edge which is an optical fiber cable connecting $v_i$ and $v_j$ for transmitting qubits. In each cable, there are several cores. Each core can be used as a quantum link for the entanglement of a pair of qubits.  Therefore, multiple qubits can be assigned at one edge for the entanglement at the same time. 
We assume that the optical fiber cable contains enough cores for the entanglement between switches. 
The length of $e_{ij}$ is denoted as $L_{ij}$. 
The success rate of each attempt to generate an entanglement over  $e_{ij}$ is $p_{ij}=e^{-\alpha L_{ij}}$, where $\alpha$
is a positive constant depending on the physical material. Since $p_{ij}$ only depends on the link length and link material, successful entanglement rates for different pairs of qubits over different cores at the same edge are the same. 
If a pair of qubits  from a quantum-user pair successfully generate the entanglement, there will be a quantum channel between qubits. Each channel can transmit an ebit at each time. 

\subsection{Routing  Matrices}
We use the expected throughput of a path as a routing matrix to evaluate the performance of the quantum network. 

For a quantum-user pair $\langle s, d\rangle$, let $\mathcal{A}$ denote the set of all paths between $s$ and $d$. 
Fix a path $A \in \mathcal{A}$, where $A=\{v_0, v_{1}, v_{2}, \cdots, v_{l-1}, v_l \}$, where $v_0=s$, $v_l=d$, and $l$ denotes the distance of $A$, i.e., the number of its edges. The nodes in $A$ are listed as the order in path from the source $s$ to the destination $d$, and the adjacent nodes are connected by one quantum link.  
Every switch in path $A$ assigns $Q^A$ qubits for the entanglement, which implies the number of parallel quantum channels in path $A$ can be up to $\frac{Q^A}{2}$.

From Section~\ref{sec:ep}, to build a quantum channel successful for a quantum-user pair along a path requires all links to generate entanglement and switches to swap successfully during the fixed time period. 
The probability of one attempt to generate the entanglement successfully of all links in a quantum channel at the same time is the product of the successful entanglement rate of every single link in the channel, i.e.,  $\Pi_{i=0}^{l}p_{i(i+1)}$. 
The probability of one attempt to swap successfully in all switches of a channel at the same time is the product of every switch's successful swapping rate in the channel, i.e., $q^{l-1}$. 
Then, the successful probability to build a quantum channel for the entanglement is $\Pi_{i=0}^{l}p_{i(i+1)} q^{l-1}$. 
Formally, the routing matrices are defined as the expected throughput of path $A$ with $\frac{Q^A}{2}$ quantum channels for the quantum-user  pair $\langle s, d\rangle$: 
\begin{align}\label{eq: P}
    P=\frac{Q^A}{2} \cdot \Pi_{i=0}^{l}p_{i(i+1)} \cdot q^{l-1} 
     =\frac{Q^A}{2} e^{-\alpha \sum_{i=0}^{l}{L_{i, i+1}}} q^{l-1},
\end{align}
which indicates the expected number of ebits can be transmitted from the source to the destination in the fixed time period.
In the current setting, the routing matrices also correspond to the expected number of entangled pairs of qubits along the path that are successfully established during the fixed time period. 

\subsection{Problem Formulation}
We divide our objectives into two steps, named as \textsc{STEP I} and \textsc{STEP II}. 
In \textsc{STEP I}, our goal is to maximize the number of quantum-user pairs that can be served by the network, and a main routing path is selected for every chosen quantum-user pair.  
In \textsc{STEP II}, we aim to maximize the expected throughput of all selected quantum-user pairs from \textsc{STEP I}. The main paths selected from \textsc{STEP I} are kept to ensure there is at least one routing path for each quantum-user pair selected in \textsc{STEP I}. 

\textbf{\textsc{STEP I}}:
We first formulate the problem of \textsc{STEP I}. 
To maximize the number of quantum-user pairs, we assume that  $Q^A=1$ for any path $A \in  \mathcal{A}$, and at most one path can be selected for  a quantum-user pair. 
Let the binary variable $x_{A_m} \in \{0,1\}$ denote whether the path $A_m$ of $\langle s_m, d_m\rangle$ is chosen to be entangled in the network or not.  
The formulation to maximize the number of quantum-user pairs is as follows:
\begin{align} 
    & {\rm\bf Problem} \; \textbf{S}_1 : \max \sum_{m \in \mathcal{M}} x_{A_m}, \label{obj:pairs} \\
    &{{\rm subject \;to\;}}: \nonumber \\
     &x_{A_m} \in \{0,1\}, \ \forall m \in \mathcal{M} \label{st:x} \\
    & \sum_{A_m \in \mathcal{A}_m}x_{A_m}\leq 1, \  \forall m \in \mathcal{M} \label{st:x-1}\\
    & \sum_{m \in \mathcal{M}, x_{A_m} = 1} |v_i \cap  (A_m x_{A_m})| \leq \frac{Q_i}{2},\  \forall v_i \in \mathcal{V}, \label{st:set1} \\
    & A_m\in \mathcal{A}_m, \forall m \in \mathcal{M} \label{st:path1}, 
\end{align}
where $\mathcal{A}_m$ denotes the set of all paths between $\langle s_m, d_m \rangle$, and $A_m$ is a path of $\mathcal{A}_m$.
Constraint (\ref{st:x-1}) denotes that at least one path can be selected for each quantum-user pair which can ensure the network serves quantum-user pairs as many as possible.    
Constraint (\ref{st:set1}) indicates that for any switch $v_i \in \mathcal{V}$, the total number of qubits assigned for all path through $v_i$ cannot over its capacity $Q_i$, $|\cdot|$ in (\ref{st:set1}) denotes the number of elements in the set. 

\textbf{\textsc{STEP II}}: Next, we formulate the problem in \textsc{STEP II} to maximize the expected throughput of selected quantum-user pairs from \textsc{STEP I} by determining the qubits assigned to possible paths from the path set. 
We first reserve the qubits in the network assigned for the main paths selected in \textsc{STEP II}, and maximize the expected  throughput  for quantum-user pairs from \textsc{STEP I} in the residual graph. 
Let $\hat{\mathcal{M}}$ denote the set of quantum-user pairs selected from  \textsc{STEP I}, and $\hat{M}$ denote the number of pairs in $\hat{\mathcal{M}}$. $\hat{Q}_i$ denotes the available qubits of switch $v_i$ after \textsc{STEP I}.
The formulation is as follows: 
\begin{align}
    & {\rm\bf Problem} \; \textbf{S}_2: \max_{Q^{A_{\hat{m}}} } \sum_{\hat{m}=1}^{\hat{M}} \sum_{A_{\hat{m}} \in \mathcal{A}_{\hat{m}}} P_{A_{\hat{m}}} , \\
    &{{\rm subject \;to\;}}  \nonumber \\ 
    & A_{\hat{m}} \in \mathcal{A}_{\hat{m}}, \forall \hat{m} \in \hat{\mathcal{M}}, \\
    &Q^{A_{\hat{m}}} \in \mathbb{N},\; \forall A_{\hat{m}} \in \mathcal{A}_{\hat{m}}, \;\hat{m} \in \hat{\mathcal{M}}, \\
    & 0 \leq Q^{A_{\hat{m}}} \leq \frac{\hat{Q}_i}{2} ,  \ \ \forall A_{\hat{m}} \in \mathcal{A}_m, \;\hat{m} \in \hat{\mathcal{M}}, \;\forall v_i \in \mathcal{V},\label{st:Q_a_m} \\
    & \sum_{{m} \in \mathcal{M}} Q^{A_{\hat{m}}} |v_i \cap  A_{\hat{m}} | \leq \frac{\hat{Q}_i}{2},\ \  \forall v_i \in \mathcal{V}, \label{st:path2}
\end{align}
where $P_{A_{\hat{m}}}$ are the expected throughput of path $A_{\hat{m}}$  defined in (\ref{eq: P}) and $\mathbb{N}$ denotes the set of non-negative integers. 
$Q^{A_{\hat{m}}}$ is the qubits assigned for path $A_{\hat{m}}$. 
(\ref{st:Q_a_m}) means that switch $v_i$ cannot assign the qubits to the path over its capacity. 
(\ref{st:path2}) indicates that the number of qubits in the switch is the main limitation for paths selection of quantum-user pairs. 

\section{Entanglement Routing Algorithm of \textsc{STEP I} } \label{sec:Alg}
We first propose algorithms to solve Problem $\textbf{S}_1$ in  \textsc{STEP I}, and analyze their performance and time complexity.  
There are two parts to solve Problem $\textbf{S}_1$.
First, we relax the binary variable $x_{A_m}$ from $\{0,1 \}$ to $[0,1]$. Let Problem $\hat{\textbf{S}}_1$ denote the relaxed problem which is a standard linear programming.
However, the time complexity to solve  Problem $\hat{\textbf{S}}_1$ is extremely high because of the huge size of the path set (the detailed analyses are presented in Section~\ref{sec:stepICom}). 
Hence, we construct a smaller path set that contains sufficient paths to reduce the complexity to solve Problem $\hat{\textbf{S}}_1$. 
Second, we derive the feasible integer solution from the solution of Problem $\hat{\textbf{S}}_1$. 

\subsection{Complexity}\label{sec:stepICom} \label{sec:complexity S1}
The Problem $\textbf{S}_1$ in \textsc{STEP I} is an binary multi-commodity flow problem. It has been proved that the problem is NP-Complete~\cite{even1975complexity}. 
When relaxing the binary variable to be continuous, the fractional solution can be solved by the standard Linear-Programming techniques such as simplex~\cite{chvatal1983linear}. 

However, the overhead for computing the paths set is not considered in the previous papers~\cite{chakraborty2020entanglement,zhang2021fragFmentation}.  
An inevitable prerequisite for solving  Problem $\textbf{S}_1$ is that the routing paths set $\mathcal{A}_m, \forall m \in \mathcal{M}$ should be calculated. This will add extra extremely huge computing complexity to solve Problem $\mathbf{S}_1$. 
More specifically, there could be up to $|\mathcal{E}|!$  paths between one quantum-user pair in a complete graph (the switches can be selected multiple times), where $|\mathcal{E}|$ is the number of edges in $\mathcal{G}$. 

The huge paths sets will cause great computational overhead to solve relaxed Problem $\hat{\textbf{S}}_1$ by using standard linear programming techniques not to mention the integer solution. 
The computing complexity will be unacceptable to solve  Problem $\textbf{S}_1$ directly. 



\subsection{ Problem $\hat{\textbf{S}}_1$ Solution}
As we have discussed above, using the standard linear programming technique to solve Problem $\hat{\textbf{S}}_1$ with huge paths sets will bring unacceptable complexity. 
To address this challenge, we select the shortest distance (i.e., the number of its edges) paths of quantum-user pairs as the path set instead of all possible paths. 
Choosing shortest distance paths can consume fewer resources (e.g., the qubits in switches) to satisfy more commodities. 
This is because the network sources are limited, it is preferred to choose a shorter distance path that consumes fewer resources from the feasible path set to maximize the number of quantum-user pairs in the network. More accurate proofs are shown in \cite{ford1958suggested,barnhart2000using}.



\renewcommand{\algorithmicrequire}{\textbf{Input:}}
\renewcommand{\algorithmicensure}{\textbf{Output:}}
\begin{algorithm} 
    \small
    \caption{Selective Paths Algorithm}
    \label{alg-1}
    \begin{algorithmic}[1]
        \Require
        $\mathcal{G}=(\mathcal{V},\mathcal{E}), \mathcal{S}, \mathcal{D}, \mathcal{M}$
        \Ensure $\mathcal{A'}$
        \State $\mathcal{A'}=\emptyset$
        \ForAll {$m\in\mathcal{M}$}
        \State Obtain $M^2$ shortest distance paths of the pair $\langle s_m, d_m \rangle$ by Yen's algorithm, $\mathcal{A'}_m=\{A{'}_m^{k}\},k\in [1,M^2]$
        \State $\mathcal{A'}=\mathcal{A'}\cup\mathcal{A'}_m$
        \EndFor
        \State Sort paths in $\mathcal{A'}$ by ascending order of length
        \State Remove the path with largest length in $\mathcal{A'}$ until $|\mathcal{A'}|=M^2$
        \ForAll {$m\in\mathcal{M}$}
        \State Remove the path with largest length in $\mathcal{A'}_m$ until $|\mathcal{A'}_m|=M$
        \State $\mathcal{A'}=\mathcal{A'}\cup\mathcal{A'}_m$
        \EndFor
    \end{algorithmic}
\end{algorithm}
The detailed path set selection algorithm is summarized in Algorithm~\ref{alg-1}.
We explain how it runs as follows. 
The goal is to construct a new smaller feasible set $\mathcal{A'}$ with total $M^2$ paths for Problem $\hat{\textbf{S}}_1$, and the path set for each quantum-user pair has $M$ paths.   
We first compute $M^2$ shortest distance paths by Yen's algorithm~\cite{yen1971finding} for each quantum-user pair. 
The reasons are as follows. 
First, $M^2$ is a large enough number to ensure a source-destination pair has a set with sufficient paths. Meanwhile, $M^2$ does not bring a huge impact on the time complexity to solve the problem that will be discussed in detail later. 
Second, if we find shortest distance paths of all quantum-user pairs directly instead of for the individual pair one by one, some pairs with a small number of shortest distance paths will be less likely considered. 
From the fairness aspect, we choose the path set of each quantum-user pair one by one.  

Then, we sort those $M^3$ paths by ascending order of their distance and add $M^2$ paths with the shortest distance to $\mathcal{A'}$. 
We check whether each quantum-user pair has $M$ paths. 
If not, we continue to add paths of that quantum-user pair with the shortest distance to $\mathcal{A'}$ until $\mathcal{A'}$ includes $M$ paths of each quantum-user pair. 


With a smaller path set $\mathcal{A'}$, Problem $\hat{\textbf{S}}_1$ can be solved by the standard linear programming techniques~\cite{chvatal1983linear,cohen2021solving} with the acceptable time complexity. Let $\tilde{x}=\{ \tilde{x}_{A'_m} \in [0,1]|\; \forall A'_m \in \mathcal{A'}_m, m \in \mathcal{M}\}$ denote the set of Problem $\hat{\textbf{S}}_1$ solution.    


\subsection{Integer Solution Recovery}
The solution of Problem $\hat{\textbf{S}}_1$ may be fractional which is not feasible to Problem $\textbf{S}_1$. 
Hence, we recover the feasible integer solution of Problem $\textbf{S}_1$ from $\tilde{x}$ and select one main routing path for each selected quantum-user pair. Let $x_{A'_m}^{\dagger}$ denote the recovered integer solution from Algorithm~\ref{alg-2}.   

The detailed algorithm is described in Algorithm~\ref{alg-2}. 
We first add the paths for $\tilde{x}_{A'_m}=1$. 
Then, we implement the branch and price strategy~\cite{barnhart2000using} to derive feasible integer solution. Let $x_{A'_m}^{\dagger}$ denote the temporal integer solution iterated in Algorithm~\ref{alg-2} and Algorithm~\ref{alg-3}.  

\begin{algorithm}
    \small
    \caption{\textsc{STEP I} Integer Solution Recovery Algorithm}
    \label{alg-2}
    \begin{algorithmic}[1]
        \Require LP solution to \textsc{STEP I}, $\tilde{x}_{A'_m} \forall A'_m \in \mathcal{A'}_m, m\in\mathcal{M}$
        \Ensure Integer solution to \textsc{STEP I}, $x_{A'_m}^{\dagger},\forall A'_m \in \mathcal{A'}_m, m \in \mathcal{M}$
        \State $x_{A'_m}^{\dagger}=0,\bar{x}_{A'_m}=0,\forall A'_m \in \mathcal{A'}_m, m\in\mathcal{M}$
        \State Sort $\tilde{x}_{A'_m}$ in descending order
        \ForAll {${A'_m}, m\in\mathcal{M}$}
        \If {$\tilde{x}_{A'_m}=1$}
        \State $\bar{x}_{A'_m}=1$, mark the pair $\langle s_m, d_m \rangle$
        \EndIf
        \EndFor
        \State Find the maximum $\tilde{x}_{A'_m}<1,\forall A'_m \in \mathcal{A'}_m, m\in\mathcal{M}$ that satisfies the corresponding $\langle s_m, d_m \rangle$ is not entangled
        \State Branch-and-price ($\emptyset$, $m$, $\tilde{x}_{A'_m}, \bar{x}_{A'_m}$)
    \end{algorithmic}
\end{algorithm}

\textbf{Branch and Price Algorithm}:
The basic idea of the branch and price strategy is to compare the results from different search branches.  
To accelerate the search process, we optimize the search order and cut some poorly performing branches. 
We only choose two branches that maximize the number of entanglement pairs as the integer solution instead of all branches. The detailed process is summarized in Algorithm~\ref{alg-3}. 
\begin{algorithm}
    \small
    \caption{Branch and Price Algorithm}
    \label{alg-3}
    \begin{algorithmic}[1]
        \Require Current path $A'_m=\{s_m,v_1,v_2,...,v_l\}$, $m$, $\tilde{x}_{A'_m},\bar{x}_{A'_m}$
        \Ensure $x^{\dagger}_{A'_m}$
        \State $\mathcal{A''}=\emptyset$
        \ForAll {$A'_m\in \mathcal{A'}_m$}
        \If {$A_m\cap A'_m=A'_m$ and $A'_m$ is feasible}
            \State $\mathcal{A''}=\mathcal{A''}\cup A'_m$
        \EndIf
        \EndFor
        \If {$|\mathcal{A''}|\le 1$}
            \State Mark the pair $\langle s_m, d_m \rangle$
            \If {$|\mathcal{A''}|= 1$}
                \State $\bar{x}_{A''_m}=1$, $A''_m\in\mathcal{A''}$
            \EndIf
            \State Compare $\bar{x}_{A'_m}$ and $x_{A'_m}^{\dagger}$, update $x_{A'_m}^{\dagger}$ if necessary
            \State Find the maximum $\tilde{x}_{A'_{m'}}<1,\forall A'_{m'}\in\mathcal{A'}_{m'},m'\in\mathcal{M}$ such that $A'_{m'}$ is feasible and $\langle s_{m'}, d_{m'}\rangle$ is not marked
            \State Branch-and-price ($\emptyset$, $m'$, $\tilde{x}_{A'_m}$, $\bar{x}_{A'_m}$)
            \State Unmark the pair $\langle s_m, d_m \rangle$, $\bar{x}_{A''_m}=0$
        \Else
        \State Find the minimum $i$ s.t. exist two paths $A'_m(v_i)(j),A'_m(v_i')(j)\in\mathcal{A}'$, satisfies $v_i\neq v_i'$, $v_i\in A'_m(v_i)(j)$, $v_i'\in A'_m(v_i')(j)$
        \State Choose any $A'_m\in\mathcal{A'}_m$, append $v_{l+1},...,v_{i-1}$ to $A'_m$

        \ForAll {$A'_m(v_i)(j)\in \mathcal{A}''$}
            \State $c_{v_i}=c_{v_i}+\tilde{x}'_{A'_m(v_i)(j)}$
        \EndFor
        \State {Find two maximum $c_{v_i},c_{v_i'}$}
        \State Branch-and-price($A'_m\cup v_{i}$, $m$, $\tilde{x}_{A'_m}$, $\bar{x}_{A'_m}$)
        \State Branch-and-price($A'_m\cup v_{i}'$, $m$, $\tilde{x}_{A'_m}$, $\bar{x}_{A'_m}$)
        \State Choose the better solution 
        \EndIf
    \end{algorithmic}

\end{algorithm}

To start the search process, we first sort the $\tilde{x}_{A'_m}$ in the descending order and select the quantum-user pair $\langle s_m, d_m\rangle$ with the highest $\tilde{x}_{A'_m}$ as the initial pair in the algorithm. 

Then, we search the feasible integer path for this pair $\langle s_m,d_m \rangle$.
Start from $s_m$, we search along the path until we find a branch node $v_{b_m}$, e.g. $A^1=\{s_m,v_1,v_2\}$, $A^2=\{s_m,v_1,v_3\}$, $A^3=\{s_m,v_1,v_4\}$, the branch node is $v_1$.  
With multiple paths to select, the preference is to search for the paths with the larger $x$ value to reduce branches. 
For example, the current path to the branch node $v_{b_m}$ is denoted as $A'_m=\{s_m, v_1, v_2, \cdots, v_{b_m}\}$. 
For every possible next node $v_i$ which could be added in the path, 
we append $v_i$ to $A'_m$,  and count the total $x_{A'_m}$ value that satisfies $A'_m$ is feasible. In Algorithm \ref{alg-3}, among all possible nodes could be added to $A'_m$, 
we select two nodes denoted as $v_{i}$ and $v_{i}'$ with the top two $x$ values. 
Next, we continue to build paths in the following two branches denoted as $A'_{m}(v_{i})$ and $A'_{m}(v_{i}')$, respectively. 

We repeat this process to construct the path from $s_m$ to $d_m$ until there is only one feasible path or no feasible paths exist. 
If there is only one feasible path $A'_m$, then $\bar{x}_{A'_m}=1$, otherwise $\bar{x}_{A'_m}=0$.
Then, We mark the current entanglement pair $\langle s_m,d_m \rangle$ as searched.

After traversing branches for  $\langle s_m,d_m \rangle$, we choose the next quantum-user pair with the largest $x$ value which has not been searched. 
The search process will end if there are no quantum-user pairs to be searched. 

\subsection{Performance Analyses and Discussion}
\subsubsection{Time Complexity to solve Problem $\hat{\textbf{S}}_1$ (Algorithm~\ref{alg-1})}
The size of the new selected path set  $\mathcal{A'}$ is $O(M^2)$, thus there are $M^2$ variables in the linear programming. 
The total time complexity of our Algorithm \ref{alg-1} and solving the corresponding linear programming is  $O(M^3N (N^2+N\log N)+O((M+M^2)^{2.373})=O(M^3N^3+M^{4.746})$ (when Problem $\hat{S}_1$ is solved by \cite{cohen2021solving}, the linear programming solution with lowest time complexity is $O((M+M^2)^{2.373}$ as far as we know), where $N$ is the number of switches. 

\subsubsection{Algorithm~\ref{alg-2} and Algorithm~\ref{alg-3}} Algorithm~\ref{alg-3} is a sub-function in Algorithm~\ref{alg-2}, so we analyze them together. 

Algorithm~\ref{alg-3} is a modified branch and price algorithm with the recursion, the complexity is almost impossible to track~\cite{barnhart2000using}. Hence, we only analyze the performance guarantee here. 
Let $f^*(x_{A'_m}^*)=\sum_{m \in \mathcal{M}} x_{A'_m}^*$ and $x_{A'_m}^*$  denote the optimal result and the optimal solution of Problem $\hat{\textbf{S}}_1$ with path set $\mathcal{A}'$, respectively.  
Let $f^{\dagger}(x^{\dagger}_{A'_m})=\sum_{m \in \mathcal{M}} x_{A'_m}^{\dagger}$ and  $x_{A'_m}^{\dagger}$ denote the integer result and the integer solution from the Algorithm~\ref{alg-2}, respectively. 
The relationship between $f^*(x_{A'_m}^*)$ and $f^{\dagger}(x^{\dagger}_{A'_m})$ is stated in the following theorem. 
\begin{theorem} \label{Theorem1}
Algorithm~\ref{alg-2} is an approximation algorithm to   Problem $\textbf{S}_1$ with path set $\mathcal{A}'$, and it achieves  an approximation ratio of 2, i.e., $f^*(x_{A'_m}^*) \leq 2 f^{\dagger}(x^{\dagger}_{A'_m})$. 
\end{theorem}
The detailed proof will be provided in the extended version of this paper due to the space limitation.  

\section{Expected Throughput Maximization of \textsc{STEP II} } \label{sec:alg2}
In \textsc{STEP I}, we have determined the maximum quantum-user pairs number that can be served by the network and selected one major path for each of them. We then reserve the qubits in the network assigned for the main paths.  
In \textsc{STEP II}, we aim to maximize the expected throughput of selected quantum-user pairs in \textsc{STEP I} in the residual graph by optimizing the qubits assigned to each path in Problem $\textbf{S}_2$. 
The updated formulation of Problem $\textbf{S}_2$ is,
\begin{align}
    & {\rm\bf Problem} \; \textbf{S}_3: \max_{Q^{A'_{\hat{m}}} } \sum_{\hat{m}=1}^{\hat{M}} \sum_{A'_{\hat{m}} \in \mathcal{A'}_{\hat{m}}} P_{A'_{\hat{m}}} , \nonumber \\
    &{{\rm subject \;to\;}}  \nonumber \\ 
    & A'_{\hat{m}} \in \mathcal{A'}_{\hat{m}}, \forall \hat{m} \in \hat{\mathcal{M}}, \nonumber \\
    &Q^{A'_{\hat{m}}} \in \mathbb{N},\; \forall A'_{\hat{m}} \in \mathcal{A'}_{\hat{m}}, \;\hat{m} \in \hat{\mathcal{M}}, \nonumber \\
    & 0 \leq Q^{A'_{\hat{m}}} \leq \frac{\hat{Q}^i}{2},  
    \ \ \forall A'_{\hat{m}} \in \mathcal{A'}_m, \;\hat{m} \in \hat{\mathcal{M}}, \;\forall v_i \in \mathcal{V},\label{st:Q_a_m} \nonumber \\
    & \sum_{\hat{m} \in \hat{\mathcal{M}} } Q^{A'_{\hat{m}}} |v_i \cap  A'_{\hat{m}} | \leq \frac{\hat{Q}^i}{2}, \ \  \forall v_i \in \mathcal{V}.  \nonumber
\end{align}

Problem $\textbf{S}_3$ is an integer  multi-commodity flow problem which is NP-Complete~\cite{even1975complexity}. 
The formulation of Problem $\textbf{S}_3$ is similar with Problem $\textbf{S}_1$ except constraints (\ref{st:x}) and (\ref{st:x-1}). 
Hence, we modify algorithms in \textsc{STEP I} to address the problem in \textsc{STEP II}. The path set of Problem $\textbf{S}_3$ is constrained by the newly constructed path set $\mathcal{A'}$. 

\begin{algorithm}
    \small
    \caption{\textsc{STEP II} Integer Solution Algorithm}
    \label{alg-4}
    \begin{algorithmic}[1]
        \Require LP solution of \textsc{STEP II}, $\tilde{Q}^{A'_{\hat{m}}},\forall A'_{\hat{m}}\in\mathcal{A}'_{\hat{m}}, \hat{m} \in \hat{\mathcal{M}}$
        \Ensure Integer solution to step II, $Q_{A'_{\hat{m}}}^{ \dagger}, \forall  A'_{\hat{m}}\in\mathcal{A}'_{\hat{m}}, \hat{m} \in \hat{\mathcal{M}}$
        \State $Q_{A'_{\hat{m}}}^{ \dagger}=0,\forall A'_{\hat{m}}\in\mathcal{A}_m^I, \hat{m} \in \hat{\mathcal{M}}$
        \State Sort $\tilde{Q}^{A'_{\hat{m}}}$ in descending order
        \ForAll {${A'_{\hat{m}}}\in\mathcal{M}$}
        \While {$\tilde{Q}^{A'_{\hat{m}}}>1$}
        \State $\bar{Q}^{A'_{\hat{m}}}=\bar{Q}^{A'_{\hat{m}}}+1,\tilde{Q}^{A'_{\hat{m}}}=\tilde{Q}^{A'_{\hat{m}}}-1$
        \State Remove the corresponding qubits
        \EndWhile
        \EndFor
        \State Find the maximum $\tilde{Q}^{A'_{\hat{m}}}<1,\forall A_{\hat{m}}\in\mathcal{A}_{\hat{m}},\hat{m}\in\hat{\mathcal{M}}$ that satisfies ${A'_{\hat{m}}}$ is feasible
        \State Branch-and-price2 ($\tilde{Q}^{A'_{\hat{m}}}$, $\bar{Q}^{A'_{\hat{m}}}$)
    \end{algorithmic}
\end{algorithm}
First, we relax $Q^{A'_{\hat{m}}} \in \mathbb{N}$ to be a continuous  non-negative real number. 
The relaxed problem of Problem $\textbf{S}_3$ is denoted as Problem $\hat{\textbf{S}}_3$. 
Problem $\hat{\textbf{S}}_3$ is a continuous linear programming which can be solved by the standard linear programming method~\cite{chvatal1983linear,cohen2021solving}.  
Let $\tilde{Q}^{A'_{\hat{m}}}$ denote the solution solved from Problem $\hat{\textbf{S}}_3$.
Because $\tilde{Q}^{A'_{\hat{m}}}$ could be fractional which is not feasible, 
we design an integer recovery algorithm to derive the feasible integer solution. Let $Q^{A'_{\hat{m}} \dagger}$ denote the recovered integer solution, and $\bar{Q}^{A'_{\hat{m}}}$ denote the temporary integer solution iterated in Algorithm~\ref{alg-4} and Algorithm~\ref{alg-5}. 
\begin{algorithm}
    \small
    \caption{Branch-and-price2 Algorithm}
    \label{alg-5}
    \begin{algorithmic}[1]
        \Require $\tilde{Q}^{A'_{\hat{m}}}, \bar{Q}^{A'_{\hat{m}}}$
        \Ensure $Q^{\dagger}_{A'_{\hat{m}}}$
        \State Find the maximum $\tilde{Q}^{A'_{\hat{m}}}>0$, where $A'_{\hat{m}}$ is a feasible and unmarked path
        \If {Found such $\tilde{Q}^{A'_{\hat{m}}}$}
        \State Mark the path $A'_{\hat{m}}$
        \State $\bar{Q}^{A'_{\hat{m}}}= \bar{Q}^{A'_{\hat{m}}}+1$, remove the corresponding qubits
        \State Branch-and-price2 ($\tilde{Q}^{A'_{\hat{m}}}$,$\bar{Q}^{A'_{\hat{m}}}$)
        \State $\bar{Q}^{A'_{\hat{m}}}$ $= \bar{Q}^{A'_{\hat{m}}}-1$, add the corresponding qubits
        \State Branch-and-price2 ($\tilde{Q}^{A'_{\hat{m}}}$,$\bar{Q}^{A'_{\hat{m}}}$)
        \State Unmark the path $A'_{\hat{m}}$
        \Else
        \State Compare $\bar{Q}^{A'_{\hat{m}}}$ and $Q^{\dagger}_{A'_{\hat{m}}}$, update $Q^{\dagger}_{A'_{\hat{m}}}$ if necessary
        \EndIf
    \end{algorithmic}

\end{algorithm}

Algorithm~\ref{alg-4} first determines $\tilde{Q}^{A'_{\hat{m}}}$ that is equal or greater than 1. $\bar{Q}^{A'_{\hat{m}}}$ equals to the integer part of $\tilde{Q}^{A'_{\hat{m}}}$ solved from  Problem $\hat{\textbf{S}}_3$. 
The main difference needed to be dealt in Algorithm~\ref{alg-4} compared with Algorithm~\ref{alg-2} is that the range of $\tilde{Q}^{A'_{\hat{m}}} $  is $[0, \frac{Q_i}{2}]$ instead of $[0,1]$. This indicates that $\bar{Q}^{A'_{\hat{m}}}$ can be added greater than 1 in process (5th row in Algorithm~\ref{alg-4}). 

Then, we use Branch and price algorithm to deal with the remaining fractional part in which $\tilde{Q}^{A'_{\hat{m}}} < 1$,  and try to recover the feasible integer solution from this part. The Branch and price algorithm is presented in Algorithm~\ref{alg-5}.

Algorithm~\ref{alg-5} is similar but simpler compared with Algorithm~\ref{alg-3} because Problem $\textbf{S}_3$ does not have the constraint to limit the maximum path number of one quantum-user pair. 
For each iteration, the algorithm finds a feasible path for iteration and decides whether to occupy the current path. The algorithm marks the path to avoid repeatedly accessing the same path. When there are no paths to continue the iteration, Algorithm~\ref{alg-5} updates the optimal solution. 
Hence, Algorithm~\ref{alg-4} can derive the optimal solution.
\begin{theorem} \label{Theorem2}
The output of Algorithm~\ref{alg-4} $\{Q^{A'_{\hat{m}}\dagger}, \forall A_{\hat{m}} \in \mathcal {A}_{\hat{m}}, \hat{m}\in\hat{\mathcal{M}}\}$ is the optimal solution of Problem $\textbf{S}_3$.
\end{theorem}
The proof will be provided in the extended version due to the space limitation. The optimality of Algorithm~\ref{alg-4} is guaranteed by the optimality of Algorithm~\ref{alg-5} which can be proved the contradiction. 
The performance difference between Algorithm~\ref{alg-5} and Algorithm~\ref{alg-3} is that Algorithm~\ref{alg-3} only searches two branches and Algorithm~\ref{alg-5} searches every possible branch. This explains why the output of Algorithm~\ref{alg-3} is not optimal. 

Algorithm~\ref{alg-4} can be implemented as an independent algorithm to maximize the network expected throughput without considering selected quantum-user pairs in \textsc{STEP I}. 
We conduction simulations about Algorithm~\ref{alg-4} to maximize expected throughput directly in Section~\ref{sec:simulation} and the results reveal that Algorithm~\ref{alg-4} outperforms existing works.  

\section{Simulation Results} \label{sec:simulation}
In this section, we implement the proposed algorithms on a randomly generated network topology and compare the performance from the number of quantum-user pairs served by the network and the expected throughput with existing works. 

\subsection{Network Topology}
We generate random networks without the fixed topology.
Considering the randomness of the network topology, we generate 5 random networks and take the average value  of the measured value, i.e., the expected throughput and the network served quantum-user pairs.  
The area of the quantum network is set as $10k \times 10k$ unit square, each unit could be 1 kilometer. 
The number of switches $N$ is set as 50 and the number of quantum-user pairs $M$ is set as 20 by default.
These nodes are randomly placed in the area. Figure~\ref{pic:switch} and  Figure~\ref{pic: pair} present results with varying $M$ and $N$, respectively. 
The edge generation follows the work~\cite{waxman1988routing}. The number of edges is determined by the average degree of nodes which is set as 10 by default.  
The length of each edge is at least $\leq \frac{50}{\sqrt{N}}$. The edge capacity does not have the limitation according to our assumption in the model.
Quantum-user nodes do not connect with other quantum-user nodes directly, and they are connected with switches directly. 
We vary the network average degree $D$ in Figure~\ref{pic: D-throughput}.  
The number of qubits in each switch $Q_i$ is set as 2, and we assume each quantum user has enough qubits for the entanglement. Figure~\ref{pic:qubit} tests different $Q_i$.  
The fiber link material parameter $\alpha$ is deduced by setting  the successful entanglement rate of a single link as $0.01\%$.
The successful swapping rate $q=0.9$, and we vary $q$ in Figure~\ref{pic: q-mp}.

\subsection{Algorithm Benchmarks}
Our proposed routing design consisted of \textsc{STEP I} and \textsc{STEP II} is denoted as \textsc{Multi-R}.
We compare \textsc{Multi-R} with the following algorithms and routing matrices:
\begin{itemize}
    \item \textsc{Algorithm 4}: We skip \textsc{STEP I} and implement Algorithm~\ref{alg-4} directly over the path set $\mathcal{A'}$ to maximize the network expected throughput.
    \item \textsc{FER}~\cite{zhang2021fragFmentation}: First sort quantum-user pairs as the descending order of the expected throughput, then select the pair with the largest expected throughput until no feasible paths exist. 
    \item \textsc{Q-PASS}~\cite{shi2020concurrent}: \textsc{Q-PASS} is a similar greedy algorithm with \textsc{FER} that uses  $\sum \frac{1}{p_{i(i+1)}}$ as the routing matrix, where $p_{i(i+1)}$ is the successful entanglement rate of edge $e_{i(i+1)}$. It indicates the summation of each link creation rate in a path.
    \item \textsc{Baseline-1}(\textsc{B1}): we use the number of hops of a path (i.e., $l$ in (\ref{eq: P})) as the evaluation matrices, and run the greedy selection similar with \textsc{Q-PASS}.

\end{itemize}

\subsection{Performance  Evaluation}

\begin{figure}[ht] 
\vspace{-0.3in}
    \centering
    \subfloat[]{
        \includegraphics[width=0.46\columnwidth]{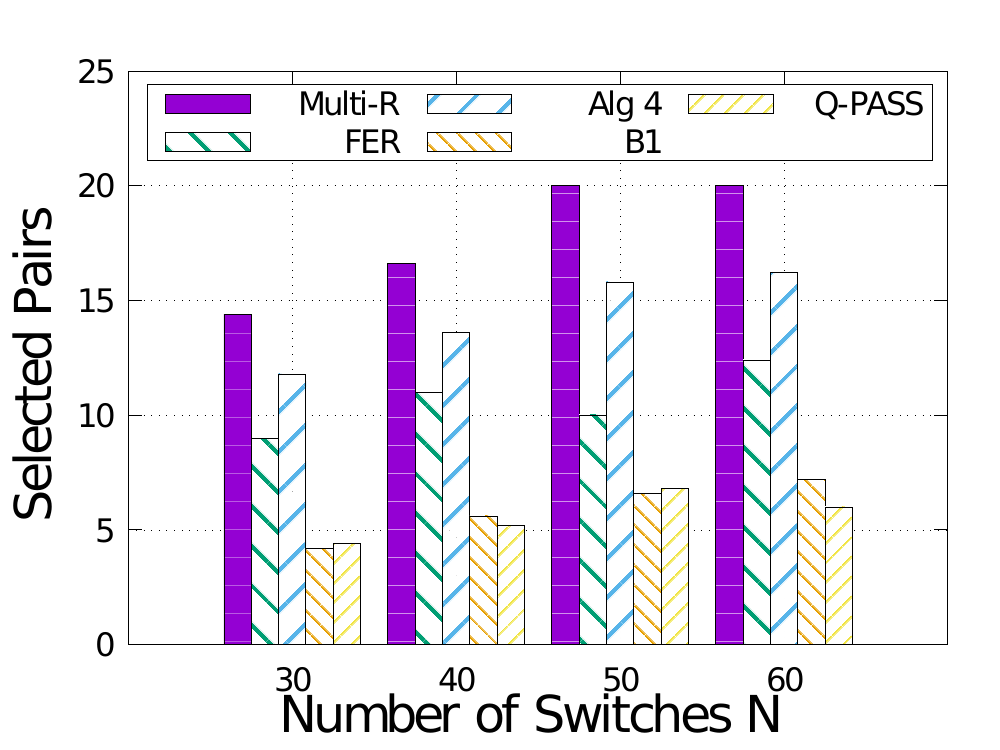} \label{pic: switch-mp}
    }
    \subfloat[]{
        \includegraphics[width=0.46\columnwidth]{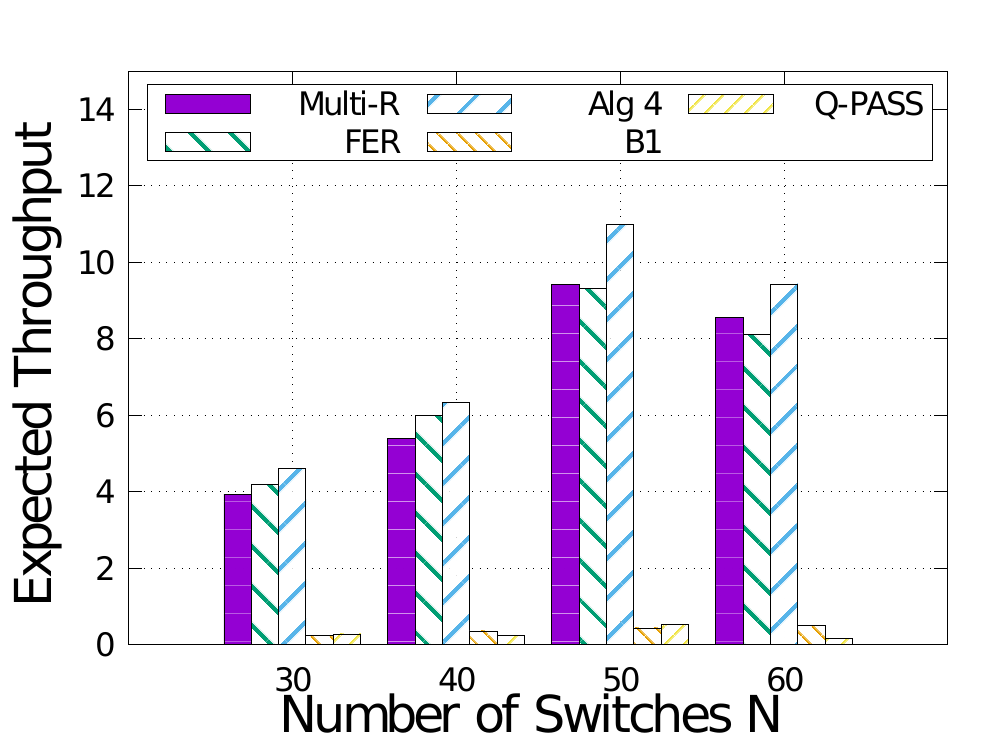}\label{pic: switch-throughput}
    }
    \caption{(a) The selected quantum-user pairs  with different  switches numbers in network.  (b) The expected throughput  with different  switches numbers in network.}
\label{pic:switch}
\vspace{-0.1in}
\end{figure}

\begin{figure}[ht] 
\vspace{-0.3in}
    \centering
    \subfloat[]{
        \includegraphics[width=0.46\columnwidth]{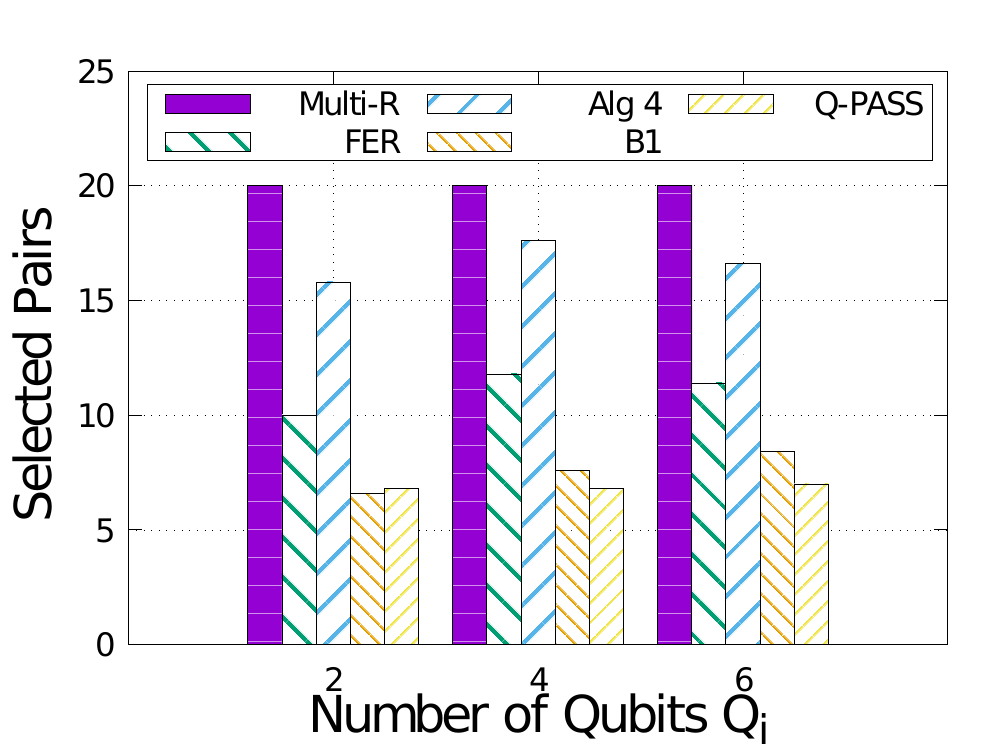} \label{pic: qubits-mp}
    }
    \subfloat[]{
        \includegraphics[width=0.46\columnwidth]{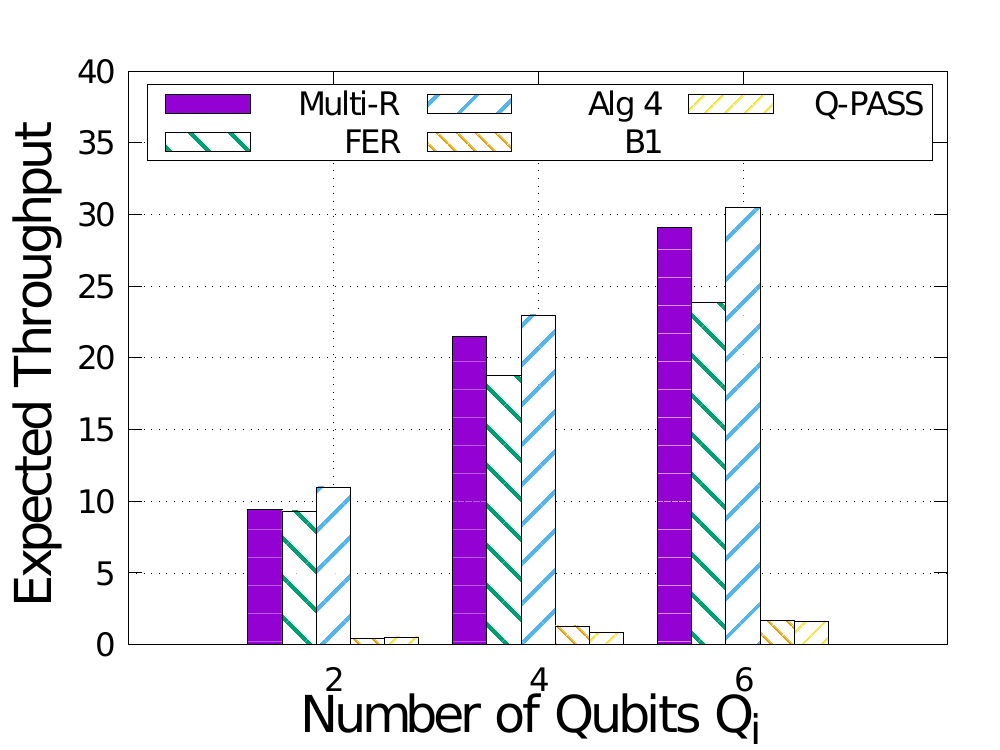}\label{pic: qubits-throughput}
    }
    \caption{(a) The selected quantum-user pairs  with varying qubits number in each switch.  (b) The expected throughput  with varying qubits number in each switch.}
\label{pic:qubit}
\vspace{-0.2
in}
\end{figure}

\begin{figure}[ht] 
\vspace{-0.3in}
    \centering
    \subfloat[]{
        \includegraphics[width=0.46\columnwidth]{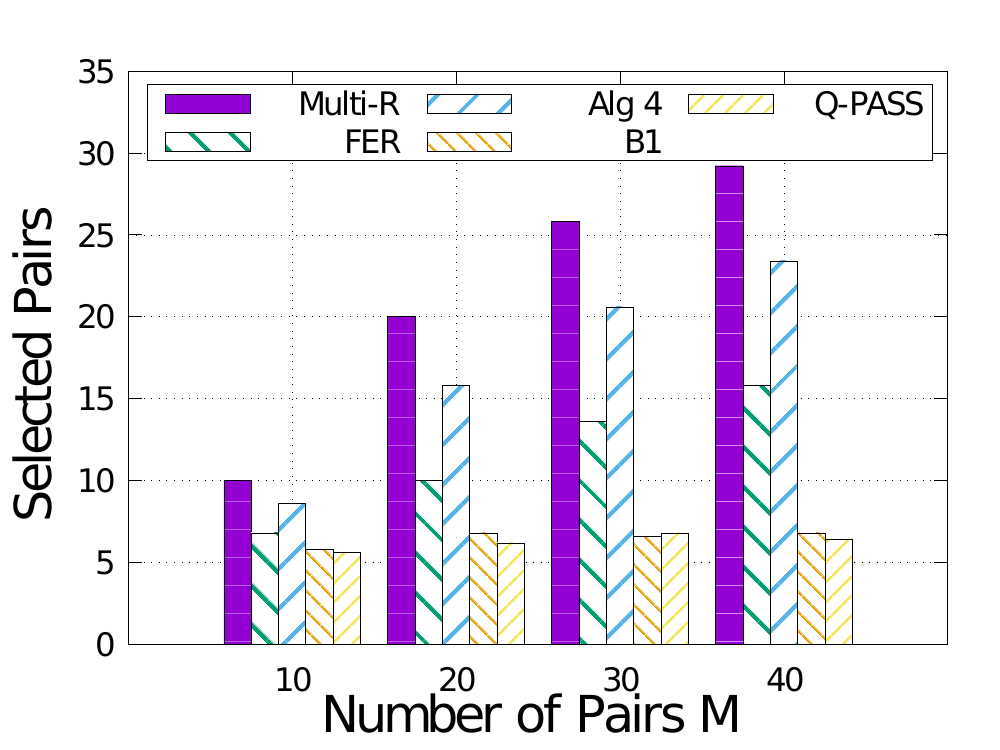} \label{pic: pair-mp}
    }
    \subfloat[]{
        \includegraphics[width=0.46\columnwidth]{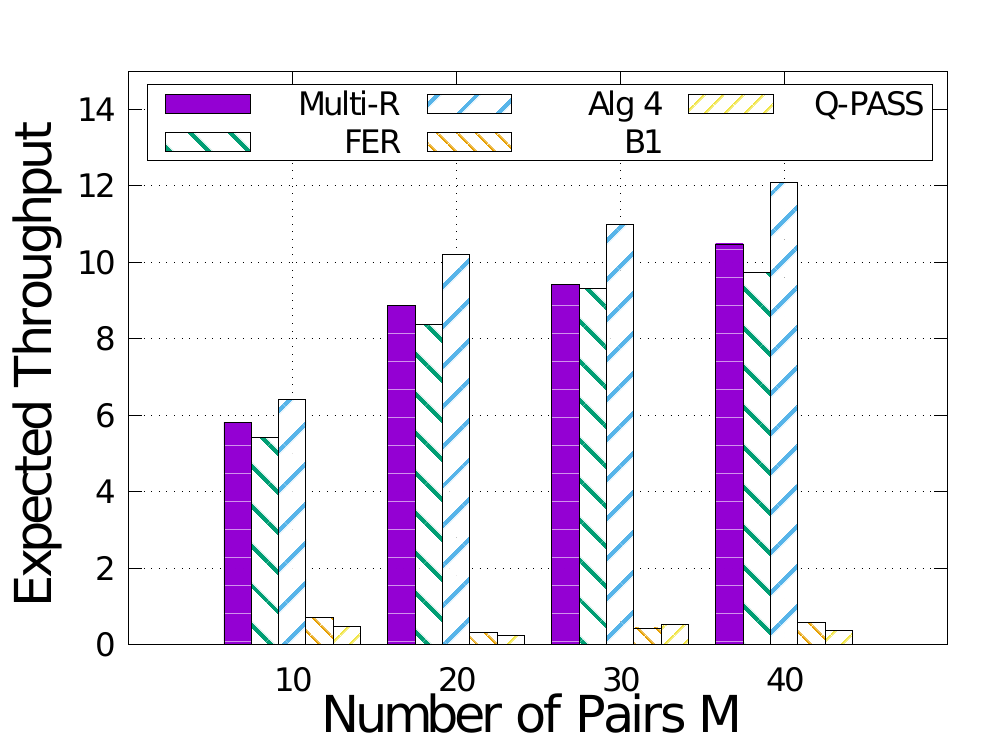}\label{pic: pair-throughput}
    }
    \caption{(a) The selected quantum-user pairs  with different quantum-user  pairs number $M$.  (b) The expected throughput  with different quantum-user pairs number $M$.}
   \vspace{-0.25in}
\label{pic: pair}
\end{figure}

\begin{figure}[ht] 
\vspace{-0.2in}
    \centering
    \subfloat[]{
        \includegraphics[width=0.46\columnwidth]{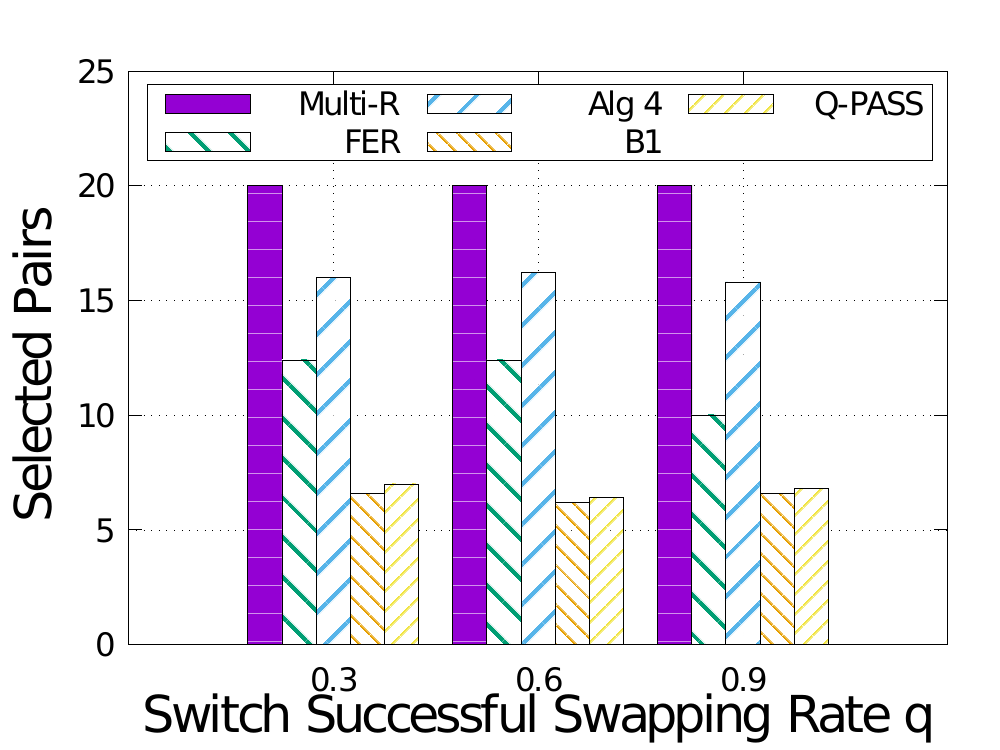} \label{pic: q-mp}
    }
    \subfloat[]{
        \includegraphics[width=0.46\columnwidth]{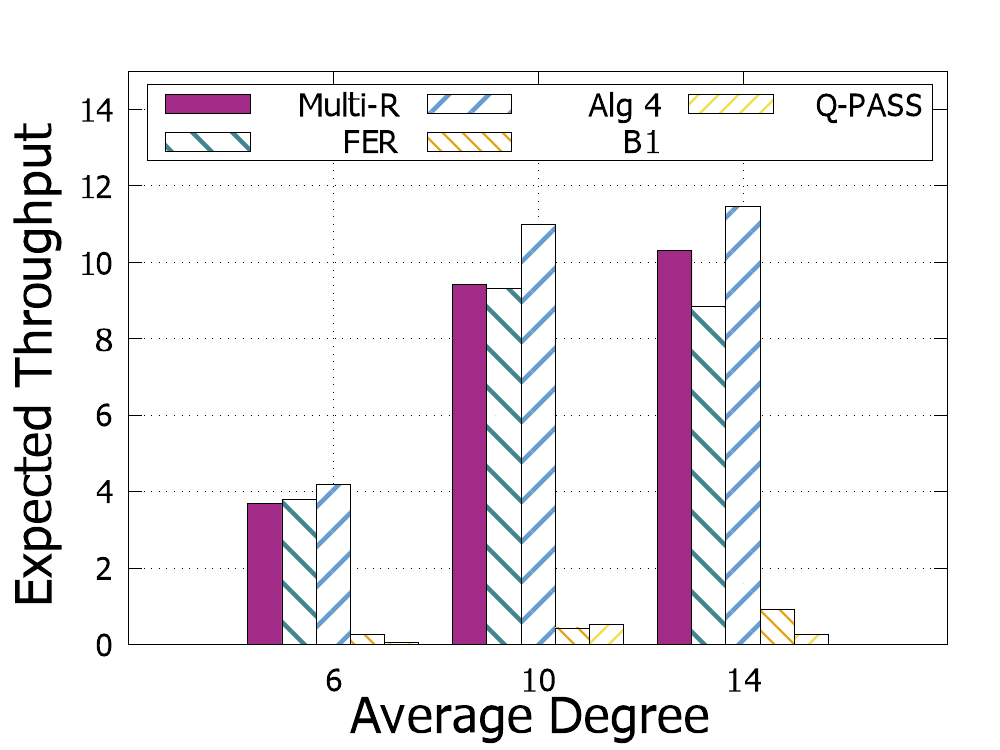}\label{pic: D-throughput}
    }
    \caption{(a) The selected quantum-user pairs  with varying successful swapping rates.  (b) The expected throughput  with varying network average degrees.}
   \vspace{-0.2in}
\end{figure}

\textbf{Number of served  quantum-user pairs.}
Figure~\ref{pic: switch-mp}, \ref{pic: qubits-mp}, \ref{pic: pair-mp}, \ref{pic: q-mp}  present the number of served quantum-user pairs with different numbers of quantum users, numbers of switches, numbers of qubits in a switch, respectively.  \textsc{Multi-R} can improve the network served quantum-user pairs number up to  85\%,  329\%,  356\%, 25\% more than  \textsc{FER}, \textsc{Q-PASS}, \textsc{B1} and \textsc{Algorithm 4}, respectively.  It reveals that our routing design to maximize the number of served quantum-user pairs can effectively let the network serve more quantum users compared with existing works. 

\textbf{Expected Throughput.} The unit of expected throughput is ebits per time slot. 
Figure~\ref{pic: switch-throughput}, \ref{pic: qubits-throughput}, \ref{pic: pair-throughput}, \ref{pic: D-throughput} compare the expected  throughput  with varying numbers of quantum users, numbers of switches, numbers of qubits in a switch. \textsc{Multi-R} has similar expected throughput compared with \textsc{FER}.
This is because we reserve main paths and  assign one qubit for each main path in \textsc{STEP I}, some of them have small expected throughput and will not be selected in \textsc{STEP II}. However, they still reserve a lot resources, which impacts the expected throughput of the network. 
Compared with \textsc{Q-PASS} and $B1$,  \textsc{Multi-R} can improve the network expected throughput up to  $16\times$ times.      
\textsc{Algorithm 4} has the largest expected throughput (improve up to 27\% compared with \textsc{FER}) among them. 
This benefits from the fact that \textsc{Algorithm 4} is  based on  optimization instead of the greedy design. Optimizing the qubits assigned to every possible path can effectively utilize the resources of the network. 

\textbf{Impact of the network parameters.}  Figure~\ref{pic:switch} and Figure~\ref{pic: D-throughput} vary  numbers of switches $N$ and  numbers of average degree $D$ of the network. 
We can observe that adding more switches and edges can improve the expected throughput and enable the network to serve more quantum-user pairs. 
When the qubits numbers of a switch is limited by the physical challenges,  increasing the density of switches and edges (optical fiber cables) can improve the network performance and serving capability. 
Figure~\ref{pic:qubit} shows the change of expected throughput and the number of served quantum-user pairs with different numbers of qubits in a switch. Even though the number of qubits in a switch is relatively small, improving the capacity of a switch can enable the network to serve more quantum-user pairs and improve their throughput. 
Figure~\ref{pic: q-mp} tests the expected  throughput with different successful swapping rate $q$. It indicates that improving the $q$ can increase the network expected  throughput.

\textsc{Multi-R} and \textsc{Algorithm 4} perform better than existing baselines in the network served quantum-user pairs number and the expected throughput, respectively, when the capacity of the network (e.g., the number of qubits of a switch, the number of switches, the number of edges) increases.  

\textbf{Impact of the quantum-user pairs number.}
We vary the number of quantum-user pairs $M$ in Figure~\ref{pic: pair}. The quantum-user pairs number denotes the network demand for the entanglement. 
When $M$ increases, the demands of the network is larger which leads to the increase of network expected throughput (shown in Figure~\ref{pic: pair-throughput}). 
\textsc{Multi-R} can let the network serve more quantum-user pairs compared with \textsc{FER}, \textsc{Q-PASS}, \textsc{B1} and \textsc{Algorithm 4}, respectively.
\textsc{Algorithm 4} has obvious improvement in expected throughput compared with existing works. 

\section{Related Work} \label{sec:related-work}
Quantum networks and their applications have drawn great attention. 
Several trials for constructing real quantum networks have been conducted, such as DARPA Quantum Network~\cite{elliott2005current}, SECOQC Vienna QKD network~\cite{peev2009secoqc}, Tokyo QKD network~\cite{sasaki2011field},  
the mobile quantum network~\cite{liu2021optical},
the integrated satellites~\cite{chen2021integrated}. 
These trial networks aim to distribute quantum keys or transmit real qubits for communication. 
However, it is still not close to be widely applied in the large-scale  quantum network in reality because of the physical and hardware limitation. 

A few studies have been conducted on the  theoretical network layer for the future large-scale quantum network. 
Numerical evaluations or simulations on the virtual  simulator are the main methods to justify the efficiency.  
Vardoyan \textit{et al.}~\cite{vardoyan2019stochastic} studied theoretical performance about the switch capacity, the memory occupancy distribution for a single switch with multiple quantum users. 
Shchhukin \textit{et al.}~\cite{shchukin2019waiting} analyzed the average waiting time for a single entanglement path based on Markov chain theory. 
Pant \textit{et al.}~\cite{pant2019routing} proposed a local routing policy for independent  switch both in single flow and multi-flow. 
Das \textit{et al.}~\cite{das2018robust} presented a routing protocol for two groups of quantum users in a Bravais lattice topology. 
Li \textit{et al.}~\cite{li2021effective} studied  the flow based network performance in a lattice network. 
Chakraborty \textit{et al.}~\cite{chakraborty2019distributed} proposed a greedy routing design in ring and grid networks. 
These papers considered the routing design in quantum networks with special topologies. These topologies may bring the advantage for the efficient design of routing protocol but they can not fit arbitrary graphs that are more common in reality. 
Shi \textit{et al.}~\cite{shi2020concurrent} proposed the routing protocol in a random graph. 
Their protocol was to add the path one by one with the largest expected throughput. 
\cite{zhang2021fragFmentation} enhanced the performance by using the remaining qubits in the network. 
However, their protocol assigned too many resources for limited quantum-user pairs that may waste the network resources and limits the number of quantum-user pairs that can be served. Their algorithms were greedy-based without considering the time complexity of choosing paths set and lacked performance guarantee. 
\cite{chakraborty2020entanglement} considered the fidelity as the main limitation for the entanglement which had high-level requirements for the capacity of the network. 

\section{Conclusion} \label{sec:conclusion}
In this paper, we have proposed an effective routing protocol for multi-entanglement routing in quantum networks to maximize the number of quantum-user pairs and their throughput at the same time. 
We have formulated our goal as two sequential integer programming steps and proposed efficient algorithms with low computational complexity  and performance guarantees. 
We have conducted simulations to show that our proposed algorithms have better performance compared with existing algorithms.

\section*{Acknowledgment}

This work is supported in part by US National Science Foundation under grant numbers 1513719, 1717731, 1730291, CNS-1717588, CNS-1730128, CNS-1919752 and an IBM Academic Award.

\bibliographystyle{./bibliography/IEEEtran}

\begin{thebibliography}{10}
\providecommand{\url}[1]{#1}
\csname url@samestyle\endcsname
\providecommand{\newblock}{\relax}
\providecommand{\bibinfo}[2]{#2}
\providecommand{\BIBentrySTDinterwordspacing}{\spaceskip=0pt\relax}
\providecommand{\BIBentryALTinterwordstretchfactor}{4}
\providecommand{\BIBentryALTinterwordspacing}{\spaceskip=\fontdimen2\font plus
\BIBentryALTinterwordstretchfactor\fontdimen3\font minus
  \fontdimen4\font\relax}
\providecommand{\BIBforeignlanguage}[2]{{%
\expandafter\ifx\csname l@#1\endcsname\relax
\typeout{** WARNING: IEEEtran.bst: No hyphenation pattern has been}%
\typeout{** loaded for the language `#1'. Using the pattern for}%
\typeout{** the default language instead.}%
\else
\language=\csname l@#1\endcsname
\fi
#2}}
\providecommand{\BIBdecl}{\relax}
\BIBdecl

\bibitem{shi2020concurrent}
S.~Shi and C.~Qian, ``Concurrent entanglement routing for quantum networks:
  Model and designs,'' in \emph{Proceedings of the Annual conference of the ACM
  Special Interest Group on Data Communication on the applications,
  technologies, architectures, and protocols for computer communication}, 2020,
  pp. 62--75.

\bibitem{garg2007faster}
N.~Garg and J.~K{\"o}nemann, ``Faster and simpler algorithms for multicommodity
  flow and other fractional packing problems,'' \emph{SIAM Journal on
  Computing}, vol.~37, no.~2, pp. 630--652, 2007.

\bibitem{pan1998experimental}
J.-W. Pan, D.~Bouwmeester, H.~Weinfurter, and A.~Zeilinger, ``Experimental
  entanglement swapping: entangling photons that never interacted,''
  \emph{Physical review letters}, vol.~80, no.~18, p. 3891, 1998.

\bibitem{zhang2019experimental}
W.-H. Zhang, G.~Chen, X.-X. Peng, X.-J. Ye, P.~Yin, X.-Y. Xu, J.-S. Xu, C.-F.
  Li, and G.-C. Guo, ``Experimental realization of robust self-testing of bell
  state measurements,'' \emph{Physical review letters}, vol. 122, no.~9, p.
  090402, 2019.

\bibitem{bouwmeester1997experimental}
D.~Bouwmeester, J.-W. Pan, K.~Mattle, M.~Eibl, H.~Weinfurter, and A.~Zeilinger,
  ``Experimental quantum teleportation,'' \emph{Nature}, vol. 390, no. 6660,
  pp. 575--579, 1997.

\bibitem{van2013designing}
R.~Van~Meter and J.~Touch, ``Designing quantum repeater networks,'' \emph{IEEE
  Communications Magazine}, vol.~51, no.~8, pp. 64--71, 2013.

\bibitem{pant2019routing}
M.~Pant, H.~Krovi, D.~Towsley, L.~Tassiulas, L.~Jiang, P.~Basu, D.~Englund, and
  S.~Guha, ``Routing entanglement in the quantum internet,'' \emph{npj Quantum
  Information}, vol.~5, no.~1, pp. 1--9, 2019.

\bibitem{nielsen2002quantum}
M.~A. Nielsen and I.~Chuang, ``Quantum computation and quantum information,''
  2002.

\bibitem{coecke2014logic}
B.~Coecke, ``The logic of entanglement,'' in \emph{Horizons of the Mind. A
  Tribute to Prakash Panangaden}.\hskip 1em plus 0.5em minus 0.4em\relax
  Springer, 2014, pp. 250--267.

\bibitem{briegel1998quantum}
H.-J. Briegel, W.~D{\"u}r, J.~I. Cirac, and P.~Zoller, ``Quantum repeaters: the
  role of imperfect local operations in quantum communication,'' \emph{Physical
  Review Letters}, vol.~81, no.~26, p. 5932, 1998.

\bibitem{bundy1984breadth}
A.~Bundy and L.~Wallen, ``Breadth-first search,'' in \emph{Catalogue of
  artificial intelligence tools}.\hskip 1em plus 0.5em minus 0.4em\relax
  Springer, 1984, pp. 13--13.

\bibitem{nain2021analysis}
P.~Nain, G.~Vardoyan, S.~Guha, and D.~Towsley, ``Analysis of a tripartite
  entanglement distribution switch,'' 2021.

\bibitem{nain2020analysis}
------, ``On the analysis of a multipartite entanglement distribution switch,''
  \emph{Proceedings of the ACM on Measurement and Analysis of Computing
  Systems}, vol.~4, no.~2, pp. 1--39, 2020.

\bibitem{shchukin2019waiting}
E.~Shchukin, F.~Schmidt, and P.~van Loock, ``Waiting time in quantum repeaters
  with probabilistic entanglement swapping,'' \emph{Physical Review A}, vol.
  100, no.~3, p. 032322, 2019.

\bibitem{zhang2021fragFmentation}
S.~Zhang, S.~Shi, C.~Qian, and K.~L. Yeung, ``Fragmentation-aware entanglement
  routing for quantum networks,'' \emph{Journal of Lightwave Technology}, 2021.

\bibitem{chakraborty2020entanglement}
K.~Chakraborty, D.~Elkouss, B.~Rijsman, and S.~Wehner, ``Entanglement
  distribution in a quantum network: A multicommodity flow-based approach,''
  \emph{IEEE Transactions on Quantum Engineering}, vol.~1, pp. 1--21, 2020.

\bibitem{kozlowski2020designing}
W.~Kozlowski, A.~Dahlberg, and S.~Wehner, ``Designing a quantum network
  protocol,'' in \emph{Proceedings of the 16th International Conference on
  emerging Networking EXperiments and Technologies}, 2020, pp. 1--16.

\bibitem{dahlberg2019link}
A.~Dahlberg, M.~Skrzypczyk, T.~Coopmans, L.~Wubben, F.~Rozp{\k{e}}dek,
  M.~Pompili, A.~Stolk, P.~Pawe{\l}czak, R.~Knegjens, J.~de~Oliveira~Filho
  \emph{et~al.}, ``A link layer protocol for quantum networks,'' in
  \emph{Proceedings of the ACM Special Interest Group on Data Communication},
  2019, pp. 159--173.

\bibitem{even1975complexity}
S.~Even, A.~Itai, and A.~Shamir, ``On the complexity of time table and
  multi-commodity flow problems,'' in \emph{16th Annual Symposium on
  Foundations of Computer Science (sfcs 1975)}.\hskip 1em plus 0.5em minus
  0.4em\relax IEEE, 1975, pp. 184--193.

\bibitem{chvatal1983linear}
V.~Chvatal, V.~Chvatal \emph{et~al.}, \emph{Linear programming}.\hskip 1em plus
  0.5em minus 0.4em\relax Macmillan, 1983.

\bibitem{ford1958suggested}
L.~R. Ford~Jr and D.~R. Fulkerson, ``A suggested computation for maximal
  multi-commodity network flows,'' \emph{Management Science}, vol.~5, no.~1,
  pp. 97--101, 1958.

\bibitem{barnhart2000using}
C.~Barnhart, C.~A. Hane, and P.~H. Vance, ``Using branch-and-price-and-cut to
  solve origin-destination integer multicommodity flow problems,''
  \emph{Operations Research}, vol.~48, no.~2, pp. 318--326, 2000.

\bibitem{yen1971finding}
J.~Y. Yen, ``Finding the k shortest loopless paths in a network,''
  \emph{management Science}, vol.~17, no.~11, pp. 712--716, 1971.

\bibitem{cohen2021solving}
M.~B. Cohen, Y.~T. Lee, and Z.~Song, ``Solving linear programs in the current
  matrix multiplication time,'' \emph{Journal of the ACM (JACM)}, vol.~68,
  no.~1, pp. 1--39, 2021.

\bibitem{van2014quantum}
R.~Van~Meter, \emph{Quantum networking}.\hskip 1em plus 0.5em minus 0.4em\relax
  John Wiley \& Sons, 2014.

\bibitem{cacciapuoti2019quantum}
A.~S. Cacciapuoti, M.~Caleffi, F.~Tafuri, F.~S. Cataliotti, S.~Gherardini, and
  G.~Bianchi, ``Quantum internet: networking challenges in distributed quantum
  computing,'' \emph{IEEE Network}, vol.~34, no.~1, pp. 137--143, 2019.

\bibitem{caleffi2018quantum}
M.~Caleffi, A.~S. Cacciapuoti, and G.~Bianchi, ``Quantum internet: from
  communication to distributed computing!'' in \emph{Proceedings of the 5th ACM
  International Conference on Nanoscale Computing and Communication}, 2018, pp.
  1--4.

\bibitem{gisin2007quantum}
N.~Gisin and R.~Thew, ``Quantum communication,'' \emph{Nature photonics},
  vol.~1, no.~3, pp. 165--171, 2007.

\bibitem{biamonte2017quantum}
J.~Biamonte, P.~Wittek, N.~Pancotti, P.~Rebentrost, N.~Wiebe, and S.~Lloyd,
  ``Quantum machine learning,'' \emph{Nature}, vol. 549, no. 7671, pp.
  195--202, 2017.

\bibitem{scarani2009security}
V.~Scarani, H.~Bechmann-Pasquinucci, N.~J. Cerf, M.~Du{\v{s}}ek,
  N.~L{\"u}tkenhaus, and M.~Peev, ``The security of practical quantum key
  distribution,'' \emph{Reviews of modern physics}, vol.~81, no.~3, p. 1301,
  2009.

\bibitem{valivarthi2020teleportation}
R.~Valivarthi, S.~I. Davis, C.~Pe{\~n}a, S.~Xie, N.~Lauk, L.~Narv{\'a}ez, J.~P.
  Allmaras, A.~D. Beyer, Y.~Gim, M.~Hussein \emph{et~al.}, ``Teleportation
  systems toward a quantum internet,'' \emph{PRX Quantum}, vol.~1, no.~2, p.
  020317, 2020.

\bibitem{chen2021integrated}
Y.-A. Chen, Q.~Zhang, T.-Y. Chen, W.-Q. Cai, S.-K. Liao, J.~Zhang, K.~Chen,
  J.~Yin, J.-G. Ren, Z.~Chen \emph{et~al.}, ``An integrated space-to-ground
  quantum communication network over 4,600 kilometres,'' \emph{Nature}, vol.
  589, no. 7841, pp. 214--219, 2021.

\bibitem{liu2021optical}
H.-Y. Liu, X.-H. Tian, C.~Gu, P.~Fan, X.~Ni, R.~Yang, J.-N. Zhang, M.~Hu,
  J.~Guo, X.~Cao \emph{et~al.}, ``Optical-relayed entanglement distribution
  using drones as mobile nodes,'' \emph{Physical Review Letters}, vol. 126,
  no.~2, p. 020503, 2021.

\bibitem{wootters2009no}
W.~K. Wootters and W.~H. Zurek, ``The no-cloning theorem,'' \emph{Physics
  Today}, vol.~62, no.~2, pp. 76--77, 2009.

\bibitem{elliott2002building}
C.~Elliott, ``Building the quantum network,'' \emph{New Journal of Physics},
  vol.~4, no.~1, p.~46, 2002.

\bibitem{li2021effective}
C.~Li, T.~Li, Y.-X. Liu, and P.~Cappellaro, ``Effective routing design for
  remote entanglement generation on quantum networks,'' \emph{npj Quantum
  Information}, vol.~7, no.~1, pp. 1--12, 2021.

\bibitem{elliott2005current}
C.~Elliott, A.~Colvin, D.~Pearson, O.~Pikalo, J.~Schlafer, and H.~Yeh,
  ``Current status of the darpa quantum network,'' in \emph{Quantum Information
  and computation III}, vol. 5815.\hskip 1em plus 0.5em minus 0.4em\relax
  International Society for Optics and Photonics, 2005, pp. 138--149.

\bibitem{peev2009secoqc}
M.~Peev, C.~Pacher, R.~All{\'e}aume, C.~Barreiro, J.~Bouda, W.~Boxleitner,
  T.~Debuisschert, E.~Diamanti, M.~Dianati, J.~Dynes \emph{et~al.}, ``The
  secoqc quantum key distribution network in vienna,'' \emph{New Journal of
  Physics}, vol.~11, no.~7, p. 075001, 2009.

\bibitem{sasaki2011field}
M.~Sasaki, M.~Fujiwara, H.~Ishizuka, W.~Klaus, K.~Wakui, M.~Takeoka, S.~Miki,
  T.~Yamashita, Z.~Wang, A.~Tanaka \emph{et~al.}, ``Field test of quantum key
  distribution in the tokyo qkd network,'' \emph{Optics express}, vol.~19,
  no.~11, pp. 10\,387--10\,409, 2011.

\bibitem{chakraborty2019distributed}
K.~Chakraborty, F.~Rozpedek, A.~Dahlberg, and S.~Wehner, ``Distributed routing
  in a quantum internet,'' \emph{arXiv preprint arXiv:1907.11630}, 2019.

\bibitem{das2018robust}
S.~Das, S.~Khatri, and J.~P. Dowling, ``Robust quantum network architectures
  and topologies for entanglement distribution,'' \emph{Physical Review A},
  vol.~97, no.~1, p. 012335, 2018.

\bibitem{vardoyan2019stochastic}
G.~Vardoyan, S.~Guha, P.~Nain, and D.~Towsley, ``On the stochastic analysis of
  a quantum entanglement switch,'' \emph{ACM SIGMETRICS Performance Evaluation
  Review}, vol.~47, no.~2, pp. 27--29, 2019.

\bibitem{waxman1988routing}
B.~M. Waxman, ``Routing of multipoint connections,'' \emph{IEEE journal on
  selected areas in communications}, vol.~6, no.~9, pp. 1617--1622, 1988.

\end{thebibliography}

\end{document}